\documentclass[pra,superscriptaddress,showpacs, twocolumn,10pt]{revtex4}
\usepackage{amssymb}
\usepackage{graphicx}

\def\>{\rangle}

\begin{document}
\newtheorem{corollary}{Corollary}
\newtheorem{definition}{Definition}
\newtheorem{example}{Example}
\newtheorem{lemma}{Lemma}
\newtheorem{proposition}{Proposition}
\newtheorem{property}{Property}
\newtheorem{theorem}{Theorem}
\newtheorem{fact}{Fact}
\renewcommand{\choose}[2]{{{#1}\atopwithdelims(){#2}}}
\title{Multiple-copy entanglement transformation and entanglement catalysis}
\author{Runyao Duan}
\email{dry02@mails.tsinghua.edu.cn}
\author{Yuan Feng}
\email{feng-y@tsinghua.edu.cn}
\author{Xin Li}
\email{x-li02@mails.tsinghua.edu.cn}
\author{Mingsheng Ying}
\email{yingmsh@tsinghua.edu.cn} \affiliation{State Key Laboratory
of Intelligent Technology and Systems, Department of Computer
Science and Technology, Tsinghua University, Beijing, China,
100084}
\date{\today}
\begin{abstract}
We prove that any multiple-copy entanglement transformation [S.
Bandyopadhyay, V. Roychowdhury, and U. Sen, Phys. Rev. A
\textbf{65}, 052315 (2002)]  can be implemented by  a suitable
entanglement-assisted local transformation [D. Jonathan and M. B.
Plenio, Phys. Rev. Lett. \textbf{83}, 3566 (1999)]. Furthermore,
we show that the combination of multiple-copy entanglement
transformation and the entanglement-assisted one is still
equivalent to the pure entanglement-assisted one.  The
mathematical structure of multiple-copy entanglement
transformations then is carefully investigated. Many interesting
properties of multiple-copy entanglement transformations are
presented, which exactly coincide with those satisfied by the
entanglement-assisted ones. Most interestingly, we show that an
arbitrarily large number of copies of state should be considered
in multiple-copy entanglement transformations.
\end{abstract}
 \maketitle

\section{Introduction}

Quantum entanglement acts as  a crucial role in the applications
of quantum information processing, such as quantum cryptography
\cite{BB84}, quantum superdense coding \cite{BS92}, and quantum
teleportation \cite{BBC+93}. It has been viewed as a new kind of
physical resource \cite{M00}. At the same time, a fruitful branch
of quantum information theory, named quantum entanglement theory,
has been developed very quickly because of the wide use of quantum
entanglement.

One of the central problems in quantum entanglement theory is to
find the conditions for an entangled state to be converted  into
another one by means of local quantum operations and classical
communication (LOCC for short). Bennett and his collaborators
\cite{BBPS96} have made significant progress in attacking this
challenging problem for the asymptotic case. While in finite
regime, the first step was made by Nielsen in Ref. \cite{NI99}
where he proved a celebrated theorem, which presents a necessary
and sufficient condition for a bipartite entangled state to be
transformed to another pure one deterministically, under the
constraint of LOCC. More precisely, let $|\psi\rangle$ and
$|\varphi\rangle$ be two bipartite entangled states. Then the
transformation of $|\psi\rangle$ to $|\varphi\rangle$ can be
achieved with certainty if and only
$\lambda_{\psi}\prec\lambda_{\varphi}$, where $\lambda_{\psi}$ and
$\lambda_{\varphi}$ denote the Schmidt coefficient vectors of
$|\psi\rangle$ and $|\varphi\rangle$, respectively. The symbol
``$\prec$" stands for ``majorization relation", which is a vast
topic in linear algebra (for details about majorization, we refer
to Refs. \cite{MO79} and \cite{AU82}). In what follows we will
identify a bipartite entangled pure state $|\psi\rangle$ by its
Schmidt coefficient vector, which is just a probability vector. We
often directly call a probability vector a ``state". This should
not cause any confusions because it is well known that any two
bipartite pure states with the same Schmidt coefficient vectors
are equivalent in the sense that they can be converted into each
other by LOCC.

It is known in linear algebra that majorization relation
``$\prec$" is not a total ordering. Thus Nielsen's result in fact
implies that there exist two incomparable entangled states $x$ and
$y$ such that neither the transformation of $x$ to  $y$ nor  the
transformation of $y$ to $x$ can be realized with certainty under
LOCC. For transformations between incomparable states, Vidal
\cite{Vidal99} generalized Nielsen's result with a probabilistic
manner and found an explicit expression of the maximal conversion
probability under LOCC.  In Ref. \cite{JP99}, Jonathan and Plenio
discovered a strange property of entanglement: sometimes, an
entangled state can help in becoming impossible entanglement
transformations into possible without being consumed at all. To be
more specific, let $x=(0.4,0.4,0.1,0.1)$ and
$y=(0.5,0.25,0.25,0)$. We know that the transformation of $x$ to
$y$ cannot be realized with certainty under LOCC. Surprisingly, if
someone lends the two parties another entangled state
$z=(0.6,0.4)$, then the transformation of $x\otimes z$ to
$y\otimes z$ can be realized with certainty because $x\otimes
z\prec y\otimes z$. The effect of the state $z$ in this
transformation is just similar to that of a catalyst in a chemical
process since it can help the entanglement transformation process
without being consumed. Thus it is termed {\it catalyst} for the
transformation of $x$ to $y$. Such a transformation that uses
intermediate entanglement without consuming it is called
``entanglement-assisted local transformation" in \cite{JP99},
abbreviated to ELOCC. This phenomenon is now widely known as
entanglement catalysis.

Bandyopadhyay $et\ al.$  found another interesting phenomenon
\cite{SRS02}: there are pairs of incomparable bipartite entangled
states that are comparable when multiple copies are provided. Take
the above $x$ and $y$ as an  example. Although $x\nprec y$ and
$x^{\otimes 2}\nprec y^{\otimes 2}$, we do have $x^{\otimes
3}\prec y^{\otimes 3}$, which means that if Alice and Bob share
three copies of source state $x$, then they can transform them
together to the same number of copies of $y$ with certainty
without any catalyst. It demonstrates  that the effect of a
catalyst can, at least in the above situation, be implemented by
preparing more copies of the original state and transforming these
copies together. This kind of transformation is called by
Bandyopadhyay $et\ al.$  ``non-asymptotic bipartite pure-state
entanglement transformation" \cite{SRS02}. More intuitively, we
call it ``multiple-copy entanglement transformation", or MLOCC for
short \cite{DF04a}. Some important aspects of MLOCC have been
investigated \cite{SRS02}.

In Ref. \cite{DF04a}, we demonstrated that multiple copies of a
bipartite entangled pure state may serve as a catalyst for certain
entanglement transformation while a single copy cannot. Such a
state is called a {\it multiple-copy catalyst} for the original
transformation. In the above example, $z'=(0.55,0.45)$ is
certainly not a catalyst for the transformation from $x$ to $y$
since $x\otimes z'\nprec y\otimes z'$. The interesting thing here
is that if Alice and Bob borrow eight copies of $z'$, then they
can transform $x$ to $y$ since $x\otimes z'^{\otimes 8} \prec
y\otimes z'^{\otimes 8}$. So $z'$ is in fact a multiple-copy
catalyst for the transformation from $x$ to $y$. Moreover, a
tradeoff between the number of the copies of catalyst and that of
the source state is observed in Ref. \cite{DF04a}: the more copies
of the catalyst that are provided, the fewer copies of source
state that are required. That is, the combination of MLOCC and
ELOCC is very useful in the case when the number of copies of
source state and that of catalyst are limited.

Due to the importance of entanglement transformation in quantum
information processing,  the mathematical structure of
entanglement catalysis has been thoroughly studied by  Daftuar and
Klimesh in Ref. \cite{DK01}.  Especially, they showed that the
dimension of catalyst is not bounded, which disproved a Nielsen's
conjecture in his lecture notes \cite{MAJ}. Furthermore, they
proved that any nonmaximally bipartite entangled pure state can
serve as a catalyst for some transformation. This gives a positive
answer to Nielsen's other conjecture \cite{MAJ}.

However, many interesting problems related to
entanglement-assisted transformation and multiple-copy
entanglement transformation are still open. One of the most
important problems is, given a pair of bipartite entangled states
$x$ and $y$, how to find a general criterion under which a
transformation from one to the other is possible under ELOCC or
MLOCC.  In other words, how to characterize the structure of
entanglement catalysis and multiple-copy entanglement
transformation? Furthermore, is there any relation between MLOCC
and ELOCC?

 One of the main goals of current paper is devoted to the
relationship between  MLOCC and ELOCC. In Sec. II, we demonstrate
that any transformation that is possible using multiple copies is
also possible for a single copy using entanglement assistance.
More precisely, if multiple copies of $x$ can be transformed to
the same number of copies of $y$, then $x$ can also be transformed
to $y$ by borrowing a suitable catalyst $z$. Intuitively, this
result means that ELOCC is at least as powerful as MLOCC. Since
our method is constructive, and in the practical use it is always
more feasible to check whether $x^{\otimes k}\prec y^{\otimes k}$
for some $k$, our result in fact  gives a sufficient condition to
decide whether there exists some appropriate catalyst for the
transformation of $x$ to $y$. We further show that the combination
of MLOCC and ELOCC is still equivalent to pure ELOCC, again by
explicit construction of the catalyst. An interesting implication
of our results is that for any two fixed positive integers $m$ and
$n$, $m$ copies of $x$ can be transformed to the same number of
copies of $y$ under ELOCC if and only if $n$ copies of $x$ can be
transformed to the same number copies of  $y$ under ELOCC.

The relation between multiple-copy entanglement transformation and
entanglement catalysis  presented in Sec. II leads  us to study
the structure of multiple-copy entanglement transformation. In
Sec. III, we carefully investigate the mathematical structure of
MLOCC. The major difficulty in studying the structure of MLOCC is
the lack of suitable mathematical tools.  To overcome this
difficulty, we first introduce some powerful lemmas. Then we are
able to  show that almost all properties of ELOCC proved in Ref.
\cite{DK01}  are also held for MLOCC.  Especially, for any state
$y$, we obtain an analytical characterization for when MLOCC is
useful in producing $y$ (i.e., there exists $x$ such that $x$ can
be transformed to $y$ by MLOCC while $x$ is incomparable to $y$).
Combining this together with a corresponding result about the
usefulness of ELOCC previously obtained in Ref. \cite{DK01}, we
establish an equivalent relation between MLOCC and ELOCC. That is
for any state $y$, MLOCC is useful in producing $y$  if and only
if ELOCC is useful in producing the same state.

We also address an interesting question about the number of copies
of  state needed in MLOCC. We show that whenever MLOCC is useful
in producing a certain target, the number of copies of state
needed in MLOCC is not bounded. That is, if MLOCC is useful in
producing $y$, then for any positive integer $k$, there exists $x$
such that multiple copies of $x$ can be transformed to  the same
number of copies of $y$ but the number of copies needed is larger
than $k$ although the dimensions of $x$ or $y$ are very small.

In Sec. IV, we try to characterize the entanglement transformation
in a different way. This is given in terms of Renyi's entropies.
We denote by $M(y)$  the set of  entangled states which can be
transformed to $y$ by MLOCC, $T(y)$ is the set of entangled states
which can be transformed into $y$ by ELOCC, $R(y)$ is the set of
entangled states whose Renyi's entropies are not less than that of
$y$. As pointed out in Sec. II, $T(y)$ is bounded by $M(y)$ from
the bottom. It is interesting that the continuous spectrum of
Renyi's entropies enables us to give a nested sets which bind
$T(y)$ from a different direction. Some interesting properties of
$R(y)$ are also discussed briefly.

In Sec. V, we draw a conclusion together with some open problems
for further study.

\section{Relation between MLOCC and ELOCC}

The purpose of this section is to examine the relationship between
multiple-copy entanglement transformation and entanglement
catalysis. Before going further, we introduce some useful
notations. Let $V^n$ denote the set of all $n$-dimensional
probability vectors. For any $x\in V^n$, the dimensionality of $x$
is often denoted by ${\rm  dim}(x)$, i.e., ${\rm dim}(x)=n$. The
notation $x^{\downarrow}$ denotes the vector obtained by sorting
the components of $x$ in nonincreasing order.  We often use
$e_l(x)$ to denote the sum of the $l$ largest components of  $x$,
i.e.,
\begin{equation}
e_l(x)=\sum_{i=1}^{l}x^{\downarrow}_i.
\end{equation}
Then the majorization  relation can be stated as
\begin{equation}\label{majdef}
x\prec y {\rm \ if\ } e_l(x)\leq e_l(y), {\rm \ for\  all\ }1\leq
l<n,
\end{equation}
where $x$ and $y$ are in $V^n$ (note  that in the case of $l=n$ it
holds equality).

With the above notations,  Nielsen's theorem can be restated as:
the transformation of $x$ to $y$ can be realized with certainty
under LOCC if and only if $x\prec y$.

Although we consider probability vectors only, sometimes  for
simplicity we omit the normalization step since the result is not
affected.  We use $x\oplus x' $ to denote the direct sum, that is,
the vector concatenating $x$ and $x'$.  Let  $A$ and $B$ be two
sets of finite dimensional vectors, then $A\oplus B$ denotes the
set of all vectors of the form $a\oplus b$ with $a\in A$ and $b\in
B$, i.e., $A\oplus B=\{a\oplus b: a\in A{\rm\ and\ }b\in B\}$.
Similarly, $A\otimes B=\{a\otimes b: a\in A {\rm\ and \ } b\in
B\}$.

In what follows, we consider deterministic transformations only
\cite{DFY04a}. For any $y\in V^n$, define
\begin{equation}\label{mlocc}
M(y)=\{x\in V^n: x^{\otimes{k}}\prec y^{{\otimes{k}}} \ {\rm for\
some}\ k\ \geq 1\}
\end{equation}
to be the set of probability vectors which, when provided with a
finite number of copies, can be transformed to the same number of
$y$ under LOCC. Moreover, we write
\begin{equation}\label{locc}
S(y)=\{x\in V^n: x \prec y\}
\end{equation}
and
\begin{equation}\label{elocc}
T(y)=\{x\in V^n: x\otimes c\prec y\otimes c\ {\rm for \ some\
vector\ } c\}.
\end{equation}
Intuitively, $S(y)$ denote all the probability vectors which can
be transformed to $y$  directly by LOCC while $T(y)$ denotes the
ones which can be transformed to $y$ by LOCC with the help of some
suitable catalyst. The latter definition is owed to Nielsen in his
lecture notes \cite{MAJ}, also see Ref. \cite{DK01}. Sometimes we
write $x \prec_T y$ if $x\in T(y)$.

Now we ask: what  is the precise relationship between
multiple-copy entanglement transformations and
entanglement-assisted ones in general?  The following theorem
gives a sharp answer to this question. It shows that for any
probability vector $y$, $M(y)$ is just a subset of $T(y)$. This
also gives a positive and stronger answer for the problem
mentioned in the conclusion part of Ref. \cite{SRS02}: if
$x^{\otimes k}$ and $y^{\otimes k}$ are comparable under LOCC for
some $k$, then for any positive integer $l$, $x^{\otimes l}$ and
$y^{\otimes l}$ are comparable under ELOCC.

Before stating the main  theorem, we present some  simple
properties of  majorization \cite{MO79}.
\begin{proposition}\label{proposition1}\upshape
Suppose that $x\prec y$ and $x'\prec y'$, then $x\oplus x'\prec
y\oplus y'$ and $x\otimes x'\prec y\otimes y'$. More compactly,
$S(y)\oplus S(y')\subseteq S(y\oplus y')$ and $S(y)\otimes
S(y')\subseteq S(y\otimes y')$.
\end{proposition}

The following theorem is one of the main results of the current
paper.
\begin{theorem}\label{mloccinelocc}\upshape
For any probability vector $y$, $M(y)\subseteq T(y)$.
\end{theorem}

\textbf{Proof.} Take $x\in M(y)$. By definition, there exists a
positive integer $k$ such that
\begin{equation}\label{xinmy}
x^{\otimes k}\prec y^{\otimes k}.
\end{equation}

We define a vector
\begin{equation}\label{catalystc}
c=x^{\otimes (k-1)}\oplus x^{\otimes (k-2)}\otimes y \oplus \cdots
\oplus x\otimes y^{\otimes (k-2)}\oplus y^{\otimes (k-1)}.
\end{equation}

Applying Proposition \ref{proposition1} and  noticing Eqs.
(\ref{xinmy}) and (\ref{catalystc}), we can easily check
\begin{equation}\label{done}
x\otimes c\prec y\otimes c.
\end{equation}
So $x\in T(y)$. That completes our proof of this theorem. \hfill
$\blacksquare$

For a positive integer $k$, we define by $M_k(y)$ the set of all
$n$-dimensional probability vectors $x$ such that $x^{\otimes k}$
is majorized by $y^{\otimes k}$. That is,
\begin{equation}\label{mky}
M_k(y)=\{x\in V^n : x^{\otimes k} \prec y^{\otimes k}\}.
\end{equation}
Similarly,
\begin{equation}\label{tky}
T_k(y)=\{x\in V^n: {\rm \ for\ some\ }c\in V^k,x\otimes c\prec
y\otimes c\}.
\end{equation}
According to the proof of  Theorem \ref{mloccinelocc}, we have the
following:

\begin{corollary}\label{mkintk} \upshape
For any $n$-dimensional probability vector $y$ and positive
integer $k$,  $M_k(y)$ is just a subset of $T_{kn^{k-1}}(y)$. That
is,  $M_k(y)\subseteq T_{kn^{k-1}}(y)$.
\end{corollary}

We have proved that every multiple-copy entanglement
transformation can be implemented by an appropriate
entanglement-assisted one. Another interesting question is whether
we can help the entanglement-assisted transformation by increasing
the number of copies of the original state. To be concise, let us
define
\begin{equation}\label{melocc}
T^M(y)=\{x\in V^n: x^{\otimes k}\otimes c\prec y^{\otimes
k}\otimes c, {\rm\ for\ some\ } k\geq 1 {\rm \ and \ } c\}.
\end{equation}
Obviously, $T(y)\subseteq T^M(y)$. But whether or not $T(y) =
T^M(y)$? The following theorem gives a positive answer to this
question.
\begin{theorem}\label{emloccinelocc}\upshape
For any $n$-dimensional probability vector $y$, $T^M(y)=T(y)$.
\end{theorem}
Intuitively, the combination of MLOCC and ELOCC is still
equivalent to pure ELOCC. This result is rather surprising since
we have demonstrated that in the situation when the resource is
limited,  the combination of ELOCC and MLOCC is strictly more
powerful than pure ELOCC \cite{DF04a}.

\textbf{Proof.} By definition, it is obvious that $ T(y)\subseteq
T^M(y)$. Suppose $x\in T^M(y)$, then there exists a positive
integer $k$ and a vector $c'$ such that
\begin{equation}\label{condition2}
x^{\otimes k}\otimes c'\prec y^{\otimes k}\otimes c'.
\end{equation}

It is a routine calculation to show that a vector $c''$ defined by
\begin{equation}\label{catalystcc}
c''=c \otimes c'
\end{equation}
is a catalyst for the transformation from $x$ to $y$, where $c$ is
defined as Eq. (\ref{catalystc}). That is,

\begin{equation}\label{done2}
x\otimes c'' \prec y\otimes c''.
\end{equation}
Thus $x\in T(y)$, and it follows that $T^M(y)\subseteq
T(y)$.\hfill $\blacksquare$\\

As a direct application of Theorem \ref{mloccinelocc},  we
reconsider an interesting phenomenon.  In Ref. \cite{LS01}, Leung
and Smolin demonstrated  that the majorization relation  is not
monotonic under tensor product, where `monotonic' means that
$x^{\otimes k}\prec y^{\otimes k}$ implies $x^{\otimes (k+1)}\prec
y^{\otimes (k+1)}$ for any $k$.  If we  extend the majorization
relation ``$\prec$" into trumping relation ``$\prec_T$", then we
may naturally hope that for any fixed positive integers $m$ and
$n$, it is held that $x^{\otimes m}\prec_T y^{\otimes m}$ if and
only if $x^{\otimes n}\prec_T y^{\otimes n}$, since both relations
can be interpreted as $x$ is more entangled than $y$. Theorem
\ref{emloccinelocc} enables us to give a rigorous proof to show
that such an interpretation is reasonable, as the following
theorem indicates:
\begin{theorem}\label{mequin}\upshape
Let  $x$ and $y$ be two probability vectors,  and let $m$ and $n$
be any two fixed positive integers. Then
\begin{equation}\label{mn}
x^{\otimes m}\prec_T y^{\otimes m} \Leftrightarrow x^{\otimes
n}\prec_T y^{\otimes n}.
\end{equation}
\end{theorem}
Intuitively, if we can transform $m$ copies of $x$ to the same
number copies of $y$ with the aid of some catalyst, then we can
also transform $n$ copies of $x$ to the same number of copies of
$y$, and vice versa.

\textbf{Proof.}  We only need to prove one direction. From
$x^{\otimes m}\prec_T y^{\otimes m}$ it follows that $x^{\otimes
m}\in T(y^{\otimes m})$ or $x\in T^M(y)$ by definition. Then $x\in
T(y)$ follows from the relation $T^M(y)= T(y)$, which also implies
$x^{\otimes n}\prec_T y^{\otimes n}$. That completes the proof. \hfill $\blacksquare$\\

With the aid of $T^M(y)=T(y)$, we have shown the equivalence of
$x^{\otimes m}\prec_T y^{\otimes m}$ and $x^{\otimes n}\prec_T
y^{\otimes n}$, which, of course, is accordant with our common
sense.

The relation $M(y)\subseteq T(y)$ has a very important
application: it provides a feasible sufficient condition  to
determine whether a given $x$ is in $T(y)$ by checking $x\in
M(y)$. In Ref. \cite{SD03}, an algorithm with the time complexity
$O(n^{2k+3.5})$ was proposed to determine whether a  given
$n$-dimensional incomparable pair $\{x, y\}$ admits a
$k$-dimensional catalyst $c$. It is a polynomial time algorithm of
$n$ when $k$ is fixed. However, in the practical use, the
dimensions of the $x$ and $y$ are fixed, while the dimension of
the potential catalyst $c$ is not fixed, i.e., $k$ is a variable.
Even for very small $n$, the above algorithm turns into an
exponential one as $k$ increases. For example, when $n=4$, the
time complexity of the algorithm is about $O(4^{2k})$. This is
intractable. On the other hand, it is easy to check that the
number of  distinct components of $x^{\otimes k}$ and $y^{\otimes
k}$ are  at most $\choose{n-1+k}{n-1}$, which is only a polynomial
of $n$ (or $k$) when $k$ (resp. $n$) is fixed. So, even when $k$
increases, we can still check the relation $x^{\otimes k}\prec
y^{\otimes k}$ efficiently. From this point of view,  it is very
important to study the structure of MLOCC carefully, which may
give us a characterization of ELOCC.
\section{Mathematical structure of multiple-copy entanglement transformation}

In the last section, the relation between multiple-copy
entanglement transformation and entanglement catalysis is
investigated. The fact that any MLOCC transformation can be
implemented by a suitable ELOCC transformation suggests that a
careful investigation of the mechanism of MLOCC is necessary.  The
aim of this section is to examine the mathematical structure
behind this mechanism. To be more specific, for any quantum state
$y$, we focus on the following three problems: (i) characterize
the interior points of $M(y)$; (ii) determine the conditions of
when MLOCC is useful in producing a given target; (iii)
demonstrate that in general, arbitrarily large number of copies of
state should be considered in MLOCC. To achieve these goals, we
first provide some basic properties of multiple-copy entanglement
transformation in Sec. IIIA. Then in Sec. IIIB, some technical
lemmas are presented. The successive three subsections consider
the above three problems, respectively.

\subsection{Some basic properties of MLOCC}

We begin with some simple properties of MLOCC, some of which have
been investigated in Ref. \cite{SRS02}.
\begin{theorem}\label{mproperty}\upshape
Let $x$ and $y$  be two $n$-dimensional probability vectors  whose
components are both arranged into non-increasing order. Then we
have that

(1) $S(y) \subseteq M(y)$.

(2) If $x\in M(y)$ then $x_{1}\leq y_{1}$ and $x_n\geq y_{n}$.

(3) If $x\in M(y)$ and $y\in M(x)$ then $x=y$. Intuitively,  $x$
and $y$ are interconvertible under MLOCC if and only if they are
equivalent up to local unitary transformations.

\end{theorem}

\textbf{Proof.} (1) is obvious from the definitions of $S(y)$ and
$M(y)$. (2) is  proved by  Lemma 1 in  Ref. \cite{SRS02}. (3)
follows immediately  from $M(y)\subseteq T(y)$ and Lemma 2 in Ref.
\cite{JP99}. \hfill$\blacksquare$

\subsection{Some technical lemmas}

Before investigating the structure  of $M(y)$ more carefully, we
need some  lemmas. For a subset $A\subseteq V^n$, the set of all
interior points of $A$ is denoted by $A^{o}$.  It is easy to see
that $x\in S^{o}(y)$ if and only if in  Eq. (\ref{majdef}), all
inequalities hold strictly and $e_n(x)=e_n(y)$.

The major difficulty in studying the structure of entanglement
catalysis and multiple-copy entanglement transformation is the
lack of suitable mathematical tools to deal with majorization
relation under tensor product. In what follows, we try to present
some useful tools to overcome this difficulty. To be more
readable, the lengthy proofs are put into the Appendix.

\begin{lemma}\label{interiorpointproduct}\upshape
If $x$ and $x'$ are interior points of  $S(y)$ and  $S(y')$,
respectively. Then $x\otimes x'$ is also an interior point of
$S(y\otimes y')$. More compactly,
\begin{equation}
S^{o}(y)\otimes S^o(y')\subseteq S^o(y\otimes y').
\end{equation}
\end{lemma}

By using Lemma \ref{interiorpointproduct} repeatedly, we have the
following:
\begin{corollary}\label{corointerior}\upshape
Let $x$, $x'$, $y$, and $y'$ as above. Suppose $k$, $p$, $q$ are
any positive integers. Then $x^{\otimes k}$ is in the interior of
$S(y^{\otimes k})$, and $x^{\otimes p}\otimes x'^{\otimes q}$ is
in the interior of $S(y^{\otimes p}\otimes y'^{\otimes q})$.
\end{corollary}

A similar result involving direct sum  is the following:
\begin{lemma}\label{interiorpointdirectsum}\upshape
If $x$ and $x'$ are interior points of $S(y)$ and $S(y')$,
respectively, then $x\oplus x'$ is still in the interior of
$S(y\oplus y')$ if and only if $y_1>y'_{n}{\rm \ and\ } y'_1>y_m$.
More compactly, we have
\begin{equation}\label{interiorsumequi}
S^{o}(y)\oplus S^{o}(y')\subseteq S^{o}(y\oplus y') {\rm \ iff\ }
y_1>y'_{n}{\rm \ and\ } y'_1>y_m.
\end{equation}
Here we assume that $y$ and $y'$ are respectively $m$-dimensional
and $n$-dimensional nonincreasingly ordered vectors. Furthermore,
we assume that $y_1>y_m$ and $y'_1>y'_n$, i.e., neither $y$ nor
$y'$ is uniform vector because otherwise the result is trivial.
\end{lemma}
Intuitively, the direct sum of $S^{o}(y)$ and $S^{o}(y')$ is still
in the interior of $S(y\oplus y')$ if and only if $y$ and $y'$
have some suitable ``overlap".

Before stating a corollary of Lemma \ref{interiorpointdirectsum},
we introduce a useful notation. We use $x^{\oplus k}$ to denote
$k$ times direct sum of $x$ itself. That is, $$ x^{\oplus
k}=\underbrace{x\oplus x\oplus \cdots \oplus x}_{k{\rm \ times\
}}.$$ Similarly, for a set $A$, $$A^{\oplus k}=\underbrace{A\oplus
A\oplus \cdots \oplus A}_{k{\rm\ times\ }}.$$

Now a direct consequence of Lemma \ref{interiorpointdirectsum} is
the following:
\begin{corollary}\label{sumrepeat}\upshape
For any probability vector $y$ and positive integer $k$, it holds
that
\begin{equation}\label{interiorsumequirepeat}
x\in S^o(y)\Leftrightarrow x^{\oplus k}\in S^o(y^{\oplus k}).
\end{equation}
\end{corollary}

Combining Lemma \ref{interiorpointdirectsum} with Corollary
\ref{sumrepeat},  we obtain the following sufficient condition for
determining  whether a given $x$ is in the interior of $S(y)$:
\begin{corollary}\label{sumrepeatset}\upshape
Suppose  that $\{(y^i)^{\oplus k_i}:1\leq i\leq m\}$ is a set of
vectors, $y^i$ is an $n_i$-dimensional vector with components in
nonincreasing order, $x^i\in S^o(y^i)$, $1\leq i\leq m$. Denote
$x=\oplus_{i=1}^m (x^i)^{\oplus k_i}$,
$y=\oplus_{i=1}^m(y^i)^{\oplus k_i}$. If (i) $y^1_1={\rm
max}\{y^i_1:1\leq i\leq m\}$, (ii) $y^m_{n_m}={\rm
min}\{y^i_{n_i}:1\leq i\leq m\}$, and (iii) $y^i_{n_i}<y^{i+1}_1
{\rm \ for \ all\ }1\leq i<m$, then $ x\in S^o(y).$
\end{corollary}
Intuitively,  if $y^1$ and $y^m$ have the maximal and the minimal
components among all the components of the set $\{(y^i)^{\oplus
k_i}:1\leq i\leq m\}$, respectively, and   the elements in the
sequence $y^1,\cdots, y^m$ overlap with each other suitably, then
$x$ is in the interior of $ S^o(y)$.

\subsection{What is the interior point of $M(y)$?}

The most basic problem about the structure of MLOCC is: given
$y\in V^n$ and $x\in M(y)$, under what conditions $x$ is an
interior point of $M(y)$? Notice that in the ELOCC case, the same
problem has been solved in Ref. \cite{DK01}. We outline the method
in Ref. \cite{DK01} as follows.

The key tool used in \cite{DK01} is a lemma connecting  $S(y)$ and
$T(y)$: if $x\in S(y)$ and $x_1<y_1$, $x_n>y_n$, then $x$ is an
interior point of $T(y)$. (Here we have assumed that both $x$ and
$y$ are in non-increasing order.)  To prove this lemma,   another
probability vector $c$ such that $x\otimes c\in S^o(y\otimes c)$
is constructed. Then  $x\in T^o(y)$ follows immediately.

Unfortunately, the method presented in \cite{DK01} cannot be
generalized to MLOCC directly although we have known
$M(y)\subseteq T(y)$. The structure of $M(y)$ seems to be much
more complicated than $T(y)$ since it should involve majorization
relation with finite times tensor product, whose property is
little known at  present.  To obtain a characterization of the
interior points  of $M(y)$, we need to obtain a similar lemma as
in \cite{DK01}. This goal can be achieved by showing that for any
$x\in S(y)$ satisfying $x_1<y_1$,  and $x_n>y_n$, $x$ is in the
interior of $M(y)$. For this purpose, we first consider a special
form of $x$. That is, $x$ is a boundary point of $S(y)$ with only
one equality $e_d(x)=e_d(y)$ ($1<d<n-1$) in the majorization
$x\prec y$. We show that in this special case, there indeed exists
$k\geq 1$ such that $x^{\otimes k}$ is an interior point of
$S^o(y^{\otimes k})$. Then we generalize the result in this
special case to a more general case, where $x$ has the form such
that $x\in S(y)$, $x_1<y_1$ and $x_n>y_n$. We surprisingly find
that for any such probability vector $x$ one can choose a suitable
positive integer $k$ such that $x$ is an interior point of
$M_k(y)$ [see Eq. (\ref{mky})], which follows that $x$ is in the
interior of $M(y)$. By Theorem \ref{mloccinelocc} we deduce that
$T(y)$ shares a similar property. Therefore our result can be
treated as an extensive generalization of Lemma 4 in Ref.
\cite{DK01}. As a direct consequence of this result, we obtain a
simple characterization of the interior points of $M(y)$.

The following lemma shows that if  $x$ is on the boundary of
$S(y)$ but with only one equality $e_d(x)=e_d(y)$ ($1<d<n-1$) in
the majorization $x\prec y$, then we can make $x^{\otimes k}$ in
the interior of $S(y^{\otimes k})$ by  choosing a suitable
positive integer $k$.
\begin{lemma}\label{uniqueinterior}\upshape
Suppose $x$ and  $y$ are in $V^n$ whose components are both in
nonincreasing order, and $d$ is a positive integer such that
$1<d<n-1$. If
\begin{equation}\label{dn}
  e_l(x)\leq e_l(y) {\ \rm\ for\ any\ }1\leq l<n,
\end{equation}
with equality if and only if $l=d$, then for positive integer $k$,
\begin{equation}\label{2interiorcondition}
x^{\otimes k}\in S^o(y^{\otimes k})\Leftrightarrow
 y_d^k<y_1^{k-1}y_{d+1} {\rm \ and\ }
                y_{d+1}^k>y_dy _n^{k-1}.
\end{equation}
\end{lemma}
The most interesting part of this lemma is that the condition on
the right-hand side of Eq. (\ref{2interiorcondition}) does not
involve $x$.

\textbf{ Proof.} Let $x'=(x_1,\ldots,x_d)$ be the vector formed by
the $d$ largest components of $x$, and $x''$ be the rest part of
$x$. $y'$ and $y''$ can be similarly defined. Then it is easy to
check that
\begin{equation}\label{2interior}
x'\in S^o(y') {\rm \ and\ } x''\in S^o(y'')
\end{equation}
by Eq. (\ref{dn}).  Also we have
\begin{equation}\label{decom2}
x=x' \oplus x''{\rm \ and\ } y=y' \oplus y''.
\end{equation}

We give a  proof of the part `$\Leftarrow$' by seeking a
sufficient condition for $x^{\otimes k}\in S^o(y^{\otimes k})$.
First we notice the following identity by binomial theorem:
\begin{equation}\label{tensorbionormial}
(y^{\otimes k})^{\downarrow}=\left(\bigoplus_{i=0}^{k}
(y'^{\otimes (k-i)}\otimes y''^{\otimes i})^{\oplus
\choose{k}{i}}\right)^{\downarrow},
\end{equation}
$x^{\otimes k}$ has a similar expression. For the sake of
convenience, we denote
\begin{equation}\label{yicon}
y^i=(y'^{\otimes (k-i)}\otimes y''^{\otimes
  i})^{\downarrow}, n_i=d^{k-i}(n-d)^i.
\end{equation}
$x^i$ has a similar meaning.

Noticing Eqs.  (\ref{2interior}) and (\ref{yicon}), we have
\begin{equation}
x^i\in S^o(y^i), {\ \ }0\leq i\leq k
\end{equation}
by Corollary \ref{corointerior}. So to ensure $x^{\otimes k}\in
S^o(y^{\otimes k})$, we only need that the set $A=\{(y^i)^{\oplus
\choose{k}{i}}:0\leq i\leq k\}$ satisfies the conditions
(i)--(iii) in Corollary \ref{sumrepeatset}. It is easy to check
that $y^0_1$ and $y^k_{n_k}$ are the maximal and the minimal
components among the components of the vectors in set $A$,
respectively. Thus the conditions (i) and (ii) are fulfilled. We
only need $A$ to satisfy the left condition (iii), i.e.,
$y^i_{n_i}<y^{i+1}_1$ for any $0\leq i<k$, or more explicitly,
\begin{equation}\label{overlapcondition}
y_d^{k-i}y_n^i<y_1^{k-(i+1)}y_{d+1}^{i+1}, {\rm \ \ }0\leq
i<k.
\end{equation}
By the monotonicity,  Eq. (\ref{overlapcondition}) is just
equivalent to the cases of $i=0$ and $i=k-1$. That is,
\begin{equation}\label{simpleoverlapcondition}
y_d^k<y_1^{k-1}y_{d+1} {\rm \ and\ } y_{d+1}^k>y_dy _n^{k-1},
\end{equation}
which is exactly the condition in  the right-hand side of Eq.
(\ref{2interiorcondition}). That completes  the proof of the part
`$\Leftarrow$'.

Now we prove the part `$\Rightarrow$'. By contradiction, suppose
the condition on the right-hand side of Eq.
(\ref{2interiorcondition}) is not satisfied. Then there should
exist $0\leq i_0<k$ that violates the conditions in Eq.
(\ref{overlapcondition}), i.e.,
\begin{equation}\label{violateoverlapcondition}
y_d^{k-i_0}y_n^{i_0}\geq y_1^{k-(i_0+1)}y_{d+1}^{i_0+1}.
\end{equation}
But then we can deduce that
\begin{equation}
\begin{array}{r}
e_{d(i_0)}(y^{\otimes
k})=\sum_{i=0}^{i_0}\choose{k}{i}e_{n_i}(y^i)
=\sum_{i=0}^{i_0}\choose{k}{i}e_{n_i}(x^i)\\
\\
\leq e_{d(i_0)}(x^{\otimes  k}),
\end{array}
\end{equation}
which contradicts  the assumption $x^{\otimes k}\in S^o(y^{\otimes
k})$, where $d(i_0)=\sum_{i=0}^{i_0}\choose{k}{i}n_i$.
That completes the proof.\hfill$\blacksquare$\\

The following lemma is an important and useful tool to prove the
properties of $M(y)$, just as Lemma 4 in Ref. \cite{DK01}.
\begin{lemma}\label{loccinteriorinmy}\upshape
Let $x$ and $y$ be two nonincreasingly sorted $n$-dimensional
probability vectors. If
\begin{equation}\label{generalizedinteriorpoint}
x\in S(y),{\ \ } x_1<y_1 {\rm \ and\ } x_n>y_n,
\end{equation}
then $x$ is in the interior of $M(y)$.
\end{lemma}

\textbf{Proof.} Let us denote $I_{x,y}$ as the set of indices
where equalities hold in $x\prec y$, i.e.,
\begin{equation}\label{equality}
I_{x,y}=\{d: e_d(x)=e_d(y), 1\leq d<n\}.
\end{equation}
If $I_{x,y}=\emptyset$, then $x\in S^o(y)$. By the relation
$S(y)\subseteq M(y)$, it follows that $x\in M^o(y)$. We only need
to consider the non-trivial case of $I_{x,y}\neq \emptyset$. In
this case $x$ is on the boundary of $S(y)$. According to Eq.
(\ref{generalizedinteriorpoint}), for any $d\in I_{x,y}$ it holds
that $1<d<n-1$.

Let  $d_1$ and $d_2$ be the minimal and the maximal elements in
$I_{x,y}$, respectively, and let $x'$ be the vector formed by the
$d_1$ largest components  and $(n-d_2)$ least components of $x$,
i.e., $x'=(x_1,\ldots, x_{d_1}, x_{d_2+1}, \ldots, x_n)$. Let
$x''$ be the rest part of $x$. $y'$, $y''$ can be similarly
defined.

From the definitions of $d_1$ and $d_2$,  we have
\begin{equation}\label{large}
x_1\geq x_{d_1}>y_{d_1}\geq y_{d_1+1}\geq x_{d_1+1}
\end{equation}
and
\begin{equation}\label{small}
x_n\leq x_{d_2+1}<y_{d_2+1}\leq y_{d_2}\leq x_{d_2}.
\end{equation}

We also have
\begin{equation}\label{submaj}
 x'\in S(y') {\rm \ and\ } x''\in S(y'').
\end{equation}

Notice that for  $x'$, it holds that $e_l(x')\leq e_l(y')$  for
any $1\leq l< {\rm dim}(x')=n-d_2+d_1$, with equality if and only
if $l=d_1$. Let us choose $k$ such that
\begin{equation}\label{d1d2cond}
y_{d_1}^{k}<y_1^{k-1}y_{d_2+1} {\rm \ and\ }
y_{d_2+1}^{k}>y_{d_1}y_n^{k-1}.
\end{equation}
By Lemma \ref{uniqueinterior}, it follows that $x'^{\otimes k }$
is an interior point of $S(y'^{\otimes k})$, i.e.,
\begin{equation}\label{interiormaj}
e_l(x'^{\otimes k})<e_l(y'^{\otimes k}), {\rm  \ for\ any\  }
1\leq l<(n-d_2+d_1)^k.
\end{equation}

Let  $b_x$ be the rest part of $x^{\otimes k}=(x'\oplus
x'')^{\otimes k}$ except the term $x'^{\otimes k}$. Then we have
\begin{equation}\label{sum}
(x^{\otimes k})^{\downarrow}=(x'^{\otimes k}\oplus
b_x)^{\downarrow}.
\end{equation}
For any $1\leq l<n^k$, according to Eq. (\ref{sum}) we rewrite
\begin{equation}\label{eq1}
e_l(x^{\otimes k})=e_{l_1}(x'^{\otimes k})+e_{l_2}(b_x),
\end{equation}
for some $0\leq l_1\leq (n-d_2+d_1)^k$, $0\leq l_2\leq
n^k-(n-d_2+d_1)^k$, and $l_1+l_2=l$. By Eq. (\ref{submaj}) and
Proposition \ref{proposition1}, it follows that
\begin{equation}\label{gsumtensor}
b_x\prec b_y,
\end{equation}
where $b_y$ is defined similar to $b_x$. So we have
\begin{equation}\label{eq2}
e_{l_1}(x'^{\otimes k})+e_{l_2}(b_x)\leq e_{l_1}(y'^{\otimes
k})+e_{l_2}(b_y).
\end{equation}
By the definition of $e_l(y^{\otimes k})$, we also have
\begin{equation}\label{eq3}
e_{l_1}(y'^{\otimes k})+e_{l_2}(b_y) \leq e_l(y^{\otimes k}).
\end{equation}
In what follows, we will prove that in Eq. (\ref{eq2}), the
inequality should be strict under the constraint of Eq.
(\ref{d1d2cond}).  Therefore, by combing Eqs.  (\ref{eq2}) and
(\ref{eq3}), we obtain
\begin{equation}\label{goalmaj}
e_l(x^{\otimes k})< e_l(y^{\otimes k}).
\end{equation}

First, we prove that if $l_1=0$ then $l_2=0$, and if $l_1={\rm
dim}(x'^{\otimes k})$ then $l_2={\rm dim}(b_x)$. In other words,
the components of $b_x$ are strictly smaller than the maximal
component of  $x'^{\otimes k}$, but strictly greater than the
minimal component of  $x'^{\otimes k}$. These two facts are
implied by Eqs.  (\ref{large}) and (\ref{small}), respectively.
Specifically,
\begin{equation}\label{large1}
{\rm max\ } x'^{\otimes k}=x_1^k>x_1^{k-1}x_{d_1+1}={\rm max\ }b_x
\end{equation}
by Eq. (\ref{large}), and
\begin{equation}\label{small1}
{\rm min\ } x'^{\otimes k}=x_n^k<x_n^{k-1}x_{d_2}={\rm min\ } b_x
\end{equation}
by Eq. (\ref{small}).

Second, we directly deduce $0<l_1<(n-d_2+d_1)^k$ from the first
step and the fact that $0<l\leq n^k$. By Eqs.  (\ref{interiormaj})
and (\ref{gsumtensor}), and the just proved fact
$0<l_1<(n-d_2+d_1)^k$,  we conclude that the inequality in Eq.
(\ref{eq2}) is strict. So we have shown that for any $1\leq
l<n^k$, Eq. (\ref{goalmaj}) holds, which  indicates that
$x^{\otimes k}$ is an interior point of $S(y^{\otimes k})$. Thus
$x$ is in the interior of $M(y)$. \hfill$\blacksquare$

For any $y\in V^n$ whose components are in nonincreasing order, we
denote by
\begin{equation}\label{gipoint}
S^{O}(y)=\{x\in V^n: x\prec y, x_1<y_1, x_n>y_n \}
\end{equation}
the  set of  generalized interior points of $S(y)$. According to
the proof of Lemma \ref{loccinteriorinmy}, we can choose a
positive integer $k$ such that $S^O(y)\subseteq [M_k(y)]^o$. To
present this result, we define
$$d_{\rm min}={\rm min}\{i:y_1>y_i\}$$ and $$d_{\rm max}={\rm
max}\{i:y_i>y_n\}.$$
Then we have the following:
\begin{corollary}\label{kmy}\upshape
For any $y\in V^n$ whose components are in nonincreasing order. If
\begin{equation}\label{dmaxdmin}
y_{d_{\rm min}}^{k}<y_1^{k-1}y_{d_{\rm max}+1} {\rm \ and\ }
y_{d_{\rm max}+1}^{k}>y_{d_{\rm min}}y_n^{k-1},
\end{equation} then $S^O(y)\subseteq [M_k(y)]^o$.
Also we have $S^O(y)\subseteq [T_{kn^{k-1}}(y)]^o$ by Corollary
\ref{mkintk}.
\end{corollary}

\textbf{Proof.} We use the same notations as Lemma
\ref{loccinteriorinmy}. Take $x\in S^O(y)$. If
$I_{x,y}=\emptyset$, then $x\in S^o(y)$, thus $x\in [M_k(y)]^o$
for any $k\geq 1$. Now assume that $I_{x,y}\neq \emptyset$. We
only need to show the fact that Eq. (\ref{dmaxdmin}) implies Eq.
(\ref{d1d2cond}).  This fact can be simply proved  as follows. By
the definitions, we have
$$d_{\rm min}\leq d_1\leq d_2\leq d_{\rm max},$$ which yields
$$y_{d_{\rm min}}\geq y_{d_1}\geq y_{d_2+1}\geq y_{d_{\rm
max}+1}.$$ Hence  Eq. (\ref{dmaxdmin}) implies Eq. (\ref{d1d2cond}). \hfill $\blacksquare$\\

Intuitively, $M(y)$ and $T(y)$ both enclose $S^O(y)$ into their
interiors when  finite copies are provided or finite dimensional
catalysts are available.

Now we can give a characterization of the interior points of
$M(y)$ as follows:
\begin{theorem}\label{myinteriorpoint}\upshape
Let $x$ and $y$ be two $n$-dimensional nonincreasing ordered
probability vectors such that  $x\in M(y)$. Then $x$ is in the
interior of $M(y)$ if and only if  $x_1<y_1$ and $x_n>y_n$.
\end{theorem}

\textbf{ Proof.}  By definition, there exists $k$ such that
$x^{\otimes k}\prec y^{\otimes k}$. Let us assume that  $x_1<y_1$
and $x_n>y_n$. Then we have $x_1^k<y_1^k$ and $x_n^k>y_n^k$. So it
follows from  Lemma \ref{loccinteriorinmy} that $x^{\otimes k}$ is
in the interior of $M(y^{\otimes k})$. Noticing that the map
$x\mapsto x^{\otimes k}$ is continuous with respect to $x$, we
deduce  that $x$ is in the interior of $\{\bar{x}:\bar{x}^{\otimes
k}\in M(y^{\otimes k})\}$, which is obviously a subset of $M(y)$.

Conversely, suppose $x$ is in the interior of $M(y)$ but $x_1\geq
y_1$ or $x_n\leq y_n$. By part (2) of Theorem \ref{mproperty}, the
only possible cases are $x_1=y_1$ or $x_n=y_n$. Then for any
$k\geq 1$, either $e_1(x^{\otimes k})=e_1(y^{\otimes k})$ or
$e_{n^k-1}(x^{\otimes k})=e_{n^k-1}(y^{\otimes k})$, both
contradicting the assumption $x\in M^o(y)$. That completes the
proof.\hfill$\blacksquare$

\subsection{When is  MLOCC  useful?}

It is desirable to know when multiple-copy entanglement
transformation has some advantage over LOCC.  When only a
three-dimensional probability vector is under consideration,  we
can simply find that $S(y)=M(y)$ since one can easily check that
for any $x,y\in V^3$ and $k\geq 1$, $x^{\otimes k}\prec y^{\otimes
k}$ is equivalent to $x\prec y$ (this result follows immediately
from part 2) of Theorem \ref{mproperty}). Thus in such a
situation, MLOCC has no advantage over LOCC, ELOCC also has no
advantageous. In Ref. \cite{DK01}, a characterization of
$T(y)=S(y)$ has been obtained. To one's surprise, $M(y)=S(y)$ has
also a simple characterization. The most interesting thing we
would like to emphasize here is that such two characterizations
are exactly the same. Thus a nice equivalent relation
$T(y)=S(y)\Leftrightarrow M(y)=S(y)$ is obtained.

The following theorem characterizes when MLOCC is more powerful
than mere LOCC:
\begin{theorem}\label{mloccuseful}\upshape
Let $y$ be an $n$-dimensional probability vector with its
components sorted nonincreasingly. Then $M(y)\neq S(y)$ if and
only if $y_1>y_l$ and $y_{l+1}>y_n$ for some $l$ such that
$1<l<n-1$.
\end{theorem}

In other words, for a state $y$, MLOCC is useful in producing $y$
if and only if $y$ has at least two successive components that are
distinct from both its smallest and largest components.

\textbf{Proof.} Suppose that there exists such $l$. Let $x$ be the
$n$-dimensional vector whose first $l$ components are each equal
to the average of the first $l$ components of $y$, and the last
$n-l$ components each equal to the average of the last $n-l$
components of $y$. More precisely, we have $x_i=e_l(y)/l$ if $i\in
\{1,\ldots, l\}$ and $x_i=[e_n(y)-e_l(y)]/(n-l)$ if $i\in
\{l+1,\ldots, n\}$. Then it is easily checked that $x \prec y$. In
fact $x$ is on the boundary of $S(y)$ since $e_l(x)=e_l(y)$.
However, by Theorem \ref{myinteriorpoint}, $x$ is in the interior
of $M(y)$; thus $M(y) \neq S(y)$.

Conversely, assume that there is no $l$ such that $1<l<n-1$, $y_1
>y_l$, and $y_{l+1}>y_n$.  Under this assumption we will prove
that for any $x\in V^n$ whose components are in nonincreasing
order, only two inequalities, namely, $x_1\leq y_1$ and $x_n\geq
y_n$, are sufficient to guarantee $x\prec y$. This together with
part (2) of Theorem \ref{mproperty} yields $M(y)\subseteq S(y)$.
For this purpose, let $d_1$ be the number of components of $y$
equal to $y_1$, and $d_2$ the number of components equal to $y_n$.
Then $x_1 \leq y_1$ indicates that $e_{j}(x)\leq e_{j}(y)$ for $j
\in \{1,\ldots,d_1\}$. Similarly, $x_n \geq y_n$ implies
$\sum_{i=j+1}^n x_i \geq \sum_{i=j+1}^n y_i$, and therefore
$e_{j}(x)\leq e_{j}(y)$, for $j \in \{n-d_2,\ldots,n-1\}$.  But
our assumption implies that $d_1+d_2 +1 \geq n$. So $e_{j}(x)\leq
e_{j}(y)$ for all $j \in \{1,\ldots,n-1\}$, and $x \prec y$. Thus
$M(y) = S(y)$.
\hfill $\blacksquare$\\

In applying the above theorem, it should be noted that the
dimension of $y$ is somewhat arbitrary, as one can append zeroes
to the vector $y$ and thereby increase its dimension without
changing the underlying quantum state.  If the nonzero components
of $y$ take exactly two distinct values, and at least two
components are equal to the smaller values of these values, then
appending zeroes will result in a vector $y'$ such that $M(y')\neq
S(y')$, although $M(y) = S(y)$. For example, $y=(0.5,0.25,0.25)$
and $y'=(0.5,0.25,0.25,0)$. The reason for this phenomenon is that
we only consider vectors $x$ with the same dimension as that of
$y$; by increasing the dimension of $y$, we increase the allowed
choices for $x$ as well. Thus, the dimension of the initial states
$x$ under consideration may determine whether $M(y) = S(y)$.

Now we can state the following weak equivalent relation between
MLOCC and ELOCC: \vspace{1em}
\begin{corollary}\label{mlocceqelocc}\upshape
Let $y$ be an $n$-dimensional probability vector. Then $M(y) =
S(y)$ if and only if $T(y) = S(y)$.
\end{corollary}
\textbf{Proof.} This is a consequence of the above theorem
and Theorem 6 in Ref. \cite{DK01}.\hfill$\blacksquare$\\

Corollary \ref{mlocceqelocc}  establishes an essential connection
between multiple-copy entanglement transformation and entanglement
catalysis. That is, for any state $y$, MLOCC is useful in
producing $y$ if and only if ELOCC is useful in producing the same
target.

\subsection{Arbitrarily large  number of copies should be considered in  MLOCC}

We will show that  for most $y$, there is no $k$ such that $M(y)=
M_k(y)$. The physical meaning of this result is that for any given
$y$, generally there does not exist an upper bound on the number
of copies of state we should provide when we try to determine
which probability vectors can be transformed to $y$ by means of
MLOCC. Our proof will proceed as follows: first we will show that
$M_k(y)$ is a closed set for any $k$, and then we will show that
$M(y)$ is in general not closed.  It then follows that $M(y) \neq
M_k(y)$.

\begin{theorem}\label{infinitecopy}\upshape
Let $y$ be an $n$-dimensional probability vector. If $M(y)\neq
S(y)$, then $M_k(y)\neq M(y)$ for any $k$.
\end{theorem}

\textbf{Proof.} We complete the proof by showing two facts: (1)
for any $k\geq 1$ and $y\in V^n$, $M_k(y)$ is closed; (2) if
$M(y)\neq S(y)$ then $M(y)$ is not closed.

First we prove fact 1). Suppose that $x^1, x^2,\ldots$ is an
arbitrary vector sequence in $M_k(y)$ that converges to $x$. By
the definition of $M_k(y)$, we have that $(x^i)^{\otimes k}\prec
y^{\otimes k}$ for each $i=1,2,\ldots$. Specifically,
$e_l[(x^i)^{\otimes k}]\leq e_l(y^{\otimes k})$ for any $1\leq
l\leq n^k$. Noticing that $e_l(x^{\otimes k})$ is continuous with
respect to $x$ when $k$ and $l$ are fixed. By taking limit
according to each $l$ we have that $e_l(x^{\otimes k})\leq
e_l(y^{\otimes k})$ for any $1\leq l\leq n^k$, which yields $x\in
M_k(y)$.

Now  we  turn to prove  fact (2).  By Theorem \ref{mloccuseful},
the assumption that $M(y)\neq S(y)$ is equivalent to the existence
of $l$ such that $1 < l < n-1$, $y_1 > y_l$ and $y_{l+1}
> y_n$. For convenience, we redefine $l$ to be the index of the
first component of $y$ that is not equal to $y_1$, and define $m$
to be the index of the last component of $y$ that is not equal to
$y_n$; clearly we  have $l < m$.  Let $\Delta = \min\{y_1-y_l,
y_m-y_n\}$ and let $x$ be the $n$-dimensional vector given by $x_l
= y_l+\Delta$, $x_m = y_m-\Delta$, and $x_i=y_i$ for $i \notin
\{l,m\}$. It is easily checked that $y \prec x$ but $x \not\prec
y$; therefore $x \notin M(y)$ by part (3) of Theorem
\ref{mproperty}. Let $w = (\frac{1}{n},\ldots,\frac{1}{n})$ and
note that $w \in S(y)$.

Suppose that $M(y)$ is a closed set. Let us consider the set
$G=\{x(t)=tx+(1-t)w:0\leq t\leq 1\}$. Obviously, $x(0)=w$ and
$x(1)=x$. Hence geometrically $G$ is just a segment connecting $w$
and $x$. Since $w$ is in the interior of $M(y)$ and $x$ is not in
$M(y)$, and moreover, $M(y)$ is closed,  $G$ should intersect
$M(y)$ at some point $x(t_0)$, where $0<t_0<1$.  That is, $x(t_0)$
should be  a boundary point of $M(y)$. However, it is easy to
check that $x(t_0)_1<y_1$ and $x(t_0)_n>y_n$. By Theorem
\ref{myinteriorpoint}, $x(t_0)$ is  an interior point of $M(y)$.
This is a contradiction. Hence $M(y)$ cannot be closed. \hfill $\blacksquare$\\

So whenever multiple-copy entanglement transformation is useful in
producing $y$ [i.e., $M(y)\neq S(y)$], an arbitrarily large number
of copies of state  must be considered.  In other words, when
$M(y) \neq S(y)$, then for any $k$ there is a $k' > k$ such that
$M_k(y)$ is a proper subset of $M_{k'}(y)$, i.e.,
$M_{k}(y)\subsetneqq M_{k'}(y)$.  An interesting question is to
ask  whether increasing the number of copies of state by $1$ will
necessarily give an improvement. That is, to decide  whether there
is any vector $y$ and $k \geq 1$ such that $M(y) \neq S(y)$ but
$M_{k+1}(y) = M_{k}(y)$.

\section{Entanglement Transformations and Renyi's Entropy}

In Sec. II, we proved that any MLOCC transformation can be
implemented by a suitable ELOCC transformation. We further  proved
that the combination of these two kind of transformations has no
advantages over pure ELOCC.  We  argued that these results in fact
give us some sufficient conditions to check whether a given
entangled state can be transformed to another one by means of
ELOCC. In this section, we tend to characterize entanglement
transformation in another way: we seek for necessary conditions of
when a given state can be transformed to another by means of
ELOCC.

We begin with a characterization of  majorization. A necessary
condition for two probability vectors $x$ and $y$ such that
$x\prec y$ is that the Shannon entropy of $x$ is not less than
that of $y$. But this is surely not a sufficient one. In fact, a
necessary and sufficient condition is given by the following lemma
\cite{HLP52}:
\begin{lemma}\label{hlp}\upshape
Let $x$ and $y$ be two $n$-dimensional vectors. Then $x\prec y$ if
and only if for any continuous concave function $f :\ \mathcal{R}
\rightarrow \mathcal{R}$,
\begin{equation}
\sum_{i=1}^{n} f(x_i) \geq \sum_{i=1}^{n} f(y_i)).
\end{equation}
\end{lemma}

Notice that the Shannon entropy $H(x)=-\sum_{i=1}^n x_i\log_2 x_i$
corresponds to  a special concave function $f(t)=-t\log_2t$. It
cannot of course sufficiently describe the relation $x\prec y$.

Renyi entropy \cite{Ren84} is a generalized version of Shannon
entropy. For any $n$-dimensional probability vector $x$ whose
components are sorted into nonincreasing order, the $\alpha$-Renyi
entropy when $\alpha \neq 1$ is defined by
\begin{equation}\label{renyientropy}
S^{(\alpha)}(x) = \frac {{\rm
sgn}(\alpha)}{1-\alpha}\log_2(\displaystyle \sum_{i=1}^{d_x}
{x_i^\alpha}).
\end{equation}
where  $d_x$ is the number of nonzero components of $x$, ${\rm
sgn}(\alpha)=1$ if $\alpha\geq 0$, otherwise ${\rm
sgn}(\alpha)=-1$. The presence of the sign function  is just for
convenience.  In Eq. (\ref{renyientropy}), we have generalized the
definition of Renyi entropy to any real number $\alpha$ although
commonly it is only defined for $\alpha\geq 0.$

Some special cases when $\alpha$ takes or tends to different
values deserve attention. First, when $\alpha$ tends to 1, the
Renyi entropy $S^{(\alpha)}(x)$ has just the Shannon entropy of
$x$ as its limit; second, when $\alpha$ tends to $+\infty$ and
$-\infty$, the Renyi entropy has limits $-\log_2 x _1$ and $\log_2
x_{d_x}$, respectively; third, when $\alpha=0$, the Renyi entropy
is just $\log_2 d_x$.  Thus it is reasonable to define that
$S^{(1)}(x)=H(x)$, $S^{(+\infty)}(x)=-\log_2 x_1$, and
$S^{(-\infty)}(x)=\log_2 x_{d_x}$.

For two $n$-dimensional probability vectors $x$ and $y$, we say
the Renyi entropy of $x$ is not less than that of $y$, if
\begin{equation}\label{eqrenyi1}
d_x>d_y {\rm \ and\ } S^{(\alpha)}(x)\geq S^{(\alpha)}(y){\rm \
for\ all\ } \alpha\geq 0,
\end{equation}
or
\begin{equation}\label{eqrenyi2}
d_x=d_y {\rm \ and\ } S^{(\alpha)}(x)\geq S^{(\alpha)}(y){\rm \
for\ all\ } \alpha\in \mathcal{R}.
\end{equation}

Let  $R(y)$ denote the set of  all $n$-dimensional probability
vectors $x$ whose Renyi entropy  is not less than that of $y$,
i.e.,
\begin{equation}\label{ry} R(y)=\{x\in V^n : x
{\rm \ satisfies\  Eqs\ (\ref{eqrenyi1})\ or\ (\ref{eqrenyi2})}\}.
\end{equation}

The following  theorem and its corollary show that the sets $T(y)$
and $M(y)$ are both contained in $R(y)$. Intuitively, if $x$ can
be transformed to $y$ by some catalyst-assisted transformation or
multiple-copy one, then the Renyi entropy of $x$ is not less than
that of $y$. \vspace{1em}
\begin{theorem}\label{tyinry}\upshape
For any $n$-dimensional probability vector $y$,  $T(y) \subseteq
R(y)$.
\end{theorem}
\textbf{Proof.} Noticing the additivity of Renyi  entropy, that
is, $S^{(\alpha)}(x\otimes c)=S^{(\alpha)}(x)+S^{(\alpha)}(c)$, we
can obtain the result of the
theorem immediately  by Lemma \ref{hlp}.\hfill$\blacksquare$\\

Combining  Theorem \ref{tyinry} with Theorem \ref{mloccinelocc} we
have the following:
\begin{corollary}\label{minry}\upshape
For any probability vector $y$, $M(y)\subseteq R(y)$.
\end{corollary}

What is very interesting here is that the fundamental properties
exposed in Theorem \ref{mproperty} that both $T(y)$ and $M(y)$
enjoy are even held for $R(y)$, just as the following theorem
shows:
\begin{theorem}\label{ryproperty}\upshape
Let $x$ and $y$ be two $n$-dimensional probability vectors whose
components are nonincreasingly ordered. Then

(1) $S(y) \subseteq R(y)$.

(2) If $x\in R(y)$ then $x_1\leq y_1$ and $x_n\geq y_n$.

(3) If $x\in R(y)$ and $y\in R(x)$ then $x = y$.

(4) If $T(y)=S(y)$ then $R(y)=S(y)$.
\end{theorem}
{\bf Proof.}  (1) follows immediately from $S(y)\subseteq T(y)$
and Theorem \ref{tyinry}.

We now prove (2). When $\alpha
> 1$, from $S^{(\alpha)}(x) \geq S^{(\alpha)}(y)$, we have $-\log_2 x_1\geq -\log_2 y_1$ by
letting $\alpha$ tend to $+\infty$. So $x_1\leq y_1$. The proof of
$x_n\geq y_n$ needs to consider the following two cases:

Case 1:  $y_n=0$, then $x_n\geq 0$ follows immediately.

Case 2: $y_n>0$. When $\alpha <0$, from $S^{(\alpha)}(x) \geq
S^{(\alpha)}(y)$, we derive $\log_2 x_n\geq \log_2 y_n$ by letting
$\alpha$ tend to $-\infty$. Thus $x_n\geq y_n$.

To prove (3), notice that when $\alpha$ ranges over positive
integer values, the equalities $S^{(\alpha)}(x)=S^{(\alpha)}(y)$
or equivalently, $\sum_i {x_i^\alpha}=\sum_i {y_i^\alpha}$ can
sufficiently force that $x = y$.

The proof of (4) is similar to Theorem \ref{mloccuseful}.
According to 1), we only need to show $R(y)\subseteq S(y)$ under
the hypothesis $T(y)=S(y)$. Take $x\in R(y)$, by 2) we have
$x_1\leq y_1$ and $x_n\geq y_n$. Then, from $T(y)=S(y)$ we deduce
that $y_l=y_1$ or $y_{l+1}=y_n$ for any $1<l<n-1$. With the same
arguments in Theorem \ref{mloccuseful}, we can prove that $x_1\leq
y_1$ and $x_n\geq y_n$  implies $x\prec y$, or equivalently, $x\in
S(y)$. Thus we complete the proof of $R(y)\subseteq S(y)$.  \hfill$\blacksquare$\\

We have shown that for any probability vector $y$, $M(y)\subseteq
T(y)\subseteq R(y)$, and they enjoy many common properties. An
interesting question that arises here is whether any pair of them
are equal? The complete answer remains open.

\section{Conclusion and Open Problems}

In conclusion,  we proved that for any probability vector $y$,
$M(y)\subseteq T(y)$. That is, any multiple-copy entanglement
transformation can be replaced by a suitable entanglement-assisted
transformation. Furthermore, we proved that $T^M(y)=T(y)$ for any
probability vector $y$, which means that the combination of the
multiple-copy entanglement transformation and the
entanglement-assisted one is also equivalent to the pure
entanglement-assisted one. Then the mathematical structure of
MLOCC has been investigated very carefully. We surprisingly found
that almost all known properties of ELOCC are also satisfied by
MLOCC.  At present, we can use $M(y)\subseteq T(y)$ and
$T^M(y)=T(y)$ as sufficient conditions to decide whether $x\in
T(y)$ by checking $x\in M(y)$ or $x\in T^M(y)$. On the other hand,
we can also use $x\notin R(y)$ to disprove that $x\in T(y)$ or
$x\in M(y)$. This method is feasible in practical use.

There are many open problems about MLOCC and ELOCC. The biggest
one is, of course, how to give a characterization of  state $y$
such that $T(y)=M(y)$.  Another interesting problem  is from the
aspect of computability: for a given state $y$,  whether it is
computable to decide a given state $x$ in $R(y)$ [$T(y)$, or
$M(y)$].

\smallskip
\textbf{Acknowledgement:} The authors wish to thank the colleagues
in the Quantum Computation and Quantum Information Research Group.
This work was partly supported by the Natural Science Foundation
of China (Grant Nos. 60273003, 60433050, 60321002, and 60305005).
Especially, Runyao Duan acknowledges the financial support of the
PhD Student Creative Foundation of Tsinghua University (Grant No.
052420003).

\smallskip\

\section*{APPENDIX}

\textbf{Proof of Lemma \ref{interiorpointproduct}:} Without loss
of generality, we assume that all the probability vectors given in
the proof are nonincreasingly ordered.  More specifically,  let
$x$ and $y$ be $m$ dimensional,  and  let $x'$ and $y'$ be $n$
dimensional. To prove our lemma, we only need to show that
\begin{equation}\label{goal}
e_l(x\otimes x')< e_l(y\otimes y'), {\rm \ for\ } 1\leq l<mn.
\end{equation}
From $x\prec y$ and $x'\prec y'$, by Proposition
\ref{proposition1} it follows that
\begin{equation}\label{tensor}
    x\otimes x'\prec x\otimes y'\prec y\otimes y'.
\end{equation}
So we have
\begin{equation}\label{majchain}
e_l(x\otimes x')=\sum_{i=1}^m x_ie_{r_i}(x')\leq \sum_{i=1}^m x_i
e_{r_i}(y')\leq e_l(y\otimes y').
\end{equation}
where $0\leq r_i\leq n$, $\sum_{i=1}^m r_i=l$. The equality is by
the definition of $e_l(x\otimes x')$; the first inequality is by
$x'\prec y'$; and the last inequality is by Eq. (\ref{tensor}). If
one of these inequalities is strict, then Eq. (\ref{goal}) holds.
We prove this by considering two cases:

Case 1. There exists an index $i_0$ such that $0< r_{i_0} < n$.
From the assumption that $x'$ is in the interior of $S(y')$, we
have
\begin{equation}\label{interiorcond}
e_l(x')< e_l(y')
 {\rm\ \ for\ all \ \ }l<n {\rm \ (especially\ for\ }
l=r_{i_{0}} {\rm\ )}.
\end{equation}
 Notice further that any component of $x$
is positive (especially $x_{i_{0}}>0$) since otherwise $x$ will
not be an interior point of $S(y)$. It follows that the first
inequality  in Eq. (\ref{majchain}) is strict.

Case 2. For any $1\leq i \leq m$, $r_i=0$ or $r_i=n$. Let $k$ be
the maximal index such that $r_k=n$. Since $l<mn$, it is easy to
show that $1\leq k<m$ and $r_i=n$ for any $1\leq i\leq k$.
Noticing $e_n(x')=e_n(y') =1 $, we have
\begin{equation}\label{strict}
e_l(x\otimes x')=\sum_{i=1}^{k} x_i e_{n}(y')< \sum_{i=1}^{k} y_i
e_n(y')\leq e_l(y\otimes y'),
\end{equation}
 where the strict inequality in Eq. (\ref{strict}) is due to
$k<n$ and the assumption that $x$ is in the interior of $S(y)$.
That completes the proof.\hfill$\blacksquare$\\

\textbf{Proof of Lemma \ref{interiorpointdirectsum}:} In the
following proof, we assume that $x\in S^o(y)$ and $x'\in S^o(y')$.

``$\Leftarrow$": Suppose that
\begin{equation}\label{overlap}
y_1>y_n' {\rm\ \ \ and\ \ } y_m<y_1'.
\end{equation} We  will prove  that $x\oplus x'$ is in the
interior of $S(y\oplus y')$. It suffices to show
\begin{equation}\label{dirmaj}
e_l(x\oplus x')<e_l(y\oplus y')
\end{equation}
for any $1\leq l< m+n$.

One can easily verify
\begin{equation}\label{chain2}
e_l(x\oplus x')=e_p(x)+e_q(x')\leq e_p(y)+e_q(y')\leq e_l(y\oplus
y'),
\end{equation}
where $0\leq p\leq m, 0\leq q\leq n$ and $p+q=l$. To complete the
proof, we  need to  consider the following two cases:

Case 1.  $0<p<m$ or $0<q<n$. By the conditions that $x\in S^o(y)$
and $x'\in S^{o}(y')$, we have
\begin{equation}
 e_p(x)<e_p(y)
{\rm \ or\ } e_q(x')<e_q(y').
\end{equation}
Then the first inequality in Eq. (\ref{chain2}) is strict, and Eq.
(\ref{dirmaj}) follows immediately.

Case 2. $p=m$, $q=0$ or $p=0$, $q=n$. They both contradict the
assumption in Eq. (\ref{overlap}). So we finish the proof of the
sufficiency part.

``$\Rightarrow$": By contradiction, suppose that Eq.
(\ref{dirmaj}) holds for very $1\leq l<m+n$ but Eq.
(\ref{overlap}) does not hold. If $y_1\leq y_n'$ then
\begin{equation}
e_n(y\oplus y')=e_n(y')=e_n(x')\leq e_n(x\oplus x'),
\end{equation}
a contradiction with Eq. (\ref{dirmaj}) when $l=n$. Similarly, if
$y_m\geq y_1'$ then
\begin{equation}
e_m(y\oplus y')=e_m(y)=e_m(x)\leq e_m(x\oplus
 x'),
\end{equation}
which contradicts  Eq. (\ref{dirmaj}) again. That completes the proof of the lemma.\hfill$\blacksquare$\\


\begin{thebibliography}{9999}

\bibitem{BB84}  C. H. Bennett and G. Brassard, Proceedings of IEEE International
Conference on Computers, Systems, and Signal Processing,
Bangalore, India, 1984, pp. 175--179.
\bibitem{BS92}  C. H. Bennett and S. J. Wiesner, Phys. Rev. Lett. 69, 2881 (1992).
\bibitem{BBC+93}C. H. Bennett, G. Brassard, C. Crepeau, R. Jozsa, A. Peres, and W.
K. Wootters, Phys. Rev. Lett. \textbf{70}, 1895 (1993).
\bibitem{M00}   M. A. Nielsen and I. L. Chuang, \textit{Quantum Computation and Quantum
Information} (Cambridge University Press, Cambridge, England
2000).
\bibitem{BBPS96}C. H. Bennett, H. J. Bernstein, S. Popescu, and B. Schumacher, Phys. Rev. A \textbf{53}, 2046 (1996).
\bibitem{NI99}  M. A. Nielsen, Phys. Rev. Lett. \textbf{83}, 436 (1999).
\bibitem{MO79}  A. W. Marshall and I. Olkin, \textit{Inequalities: Theory of Majorization and Its Applications} (Academic Press, New York, 1979).
\bibitem{AU82}  P. M. Alberti and A. Uhlmann, \textit{Stochasticity and Partial Order: Doubly
Stochastic Maps and Unitary Mixing} (Dordrecht, Boston, 1982).
\bibitem{Vidal99}G. Vidal, Phys. Rev. Lett. \textbf{83}, 1046 (1999).
\bibitem{JP99}  D. Jonathan and M. B. Plenio, Phys. Rev. Lett. \textbf{83}, 3566
(1999).
\bibitem{SRS02} S. Bandyopadhyay, V. Roychowdhury, and U. Sen, Phys. Rev. A \textbf{65},
052315 (2002).
\bibitem{DF04a} R. Y. Duan, Y. Feng, X. Li, and M. S. Ying,
quant-ph/0312010.
\bibitem{DK01}  S. Daftuar and M. Klimesh, Phys. Rev. A \textbf{64}, 042314
(2001).
\bibitem{MAJ}   M. A. Nielsen, \textit{Introduction to Majorization and Its Applications to Quantum Mechanics} (unpublished
notes), available online:
http://www.qinfo.org/talks/2002/maj/book.ps.
\bibitem{DFY04a}The underlying
mathematical structure of probabilistic entanglement
transformation is different from that of deterministic
transformations. We need to deal with a special mathematical tool
named `super majorization'. Theorem 1 and Theorem 2 can be
generalized to probabilistic entanglement tranformations  with the
aid of the properties of super majorization.
\bibitem{LS01}  D. W. Leung and J. A. Smolin, quant-ph/0103158.
\bibitem{SD03}  X. M. Sun, R. Y. Duan,  and M. S. Ying,
 IEEE Trans. Inf. Theory \textbf{51}, 75 (2005).
\bibitem{HLP52} G. H. Hardy, J. E. Littlewood, and G. Polya,
{\it Inequalities} (Cambridge University Press, Cambridge,
England, 1952).
\bibitem{Ren84} A. Renyi, \textit{Probability Theory} (North-Holland, Amsterdam,
1970).


\end{thebibliography}
\end{document}